\documentclass[12pt,a4paper]{article}
\usepackage{amsmath,amsfonts,amsthm,graphicx,lscape}
\usepackage[dvips]{psfrag}
\usepackage{cite}

\usepackage[normalem]{ulem}

\usepackage{textcomp}
\usepackage{amsmath,amsthm,amssymb,mathrsfs,cases}
\usepackage{amsfonts,graphicx,lscape}

\newcommand{\vt}[1]{\mbox{\boldmath$#1$}}
\newcommand{\sca}[2]{\langle #1, #2 \rangle}

\usepackage{accents}
\makeatletter
\def\widebar{\accentset{{\cc@style\underline{\mskip10mu}}}}
\makeatother

\numberwithin{equation}{section} 
\newtheorem{theorem}{Theorem}[section]
\newtheorem{proposition}[theorem]{Proposition}

\def\beqa{\begin{eqnarray}}
\def\enqa{\end{eqnarray}}
\def\beq{\begin{equation}}
\def\enq{\end{equation}}

\begin{document}
\title{
\vspace{-9mm}
On 
a new integrable 
generalization of the 
Toda lattice 
and 
a discrete Yajima--Oikawa system
}
\author{Takayuki \textsc{Tsuchida}
}
\maketitle
\begin{abstract} 
%
We propose a new integrable 
generalization of 
the Toda lattice 
wherein 
the original Flaschka--Manakov variables 
are coupled to 
newly introduced dependent variables; 
the general case wherein 
the additional dependent variables are 
vector-valued 
is considered.  
This 
generalization 
admits 
a Lax pair 
based on 
an extension of the Jacobi operator, 
an infinite number of 
conservation laws 
and, in a special case, 
a simple Hamiltonian structure. 
In fact, the 
second flow 
of 
this 
generalized Toda 
hierarchy 
reduces to 
the usual 
Toda lattice 
when the 
additional dependent variables 
vanish; 
the first flow of the hierarchy 
reduces to 
a long wave--short wave interaction
model, 
known as the Yajima--Oikawa system, 
in a suitable continuous limit. 
This integrable discretization of 
the Yajima--Oikawa system is essentially 
different from the 
discrete Yajima--Oikawa system proposed in 
arXiv:1509.06996 
(also see https://link.aps.org/doi/10.1103/PhysRevE.91.062902)
and studied in arXiv:1804.10224. 
%
Two 
integrable discretizations 
of the 
nonlinear Schr\"odinger 
hierarchy, 
the Ablowitz--Ladik 
hierarchy 
and the Konopelchenko--Chudnovsky hierarchy, 
are contained in 
the generalized Toda 
hierarchy 
as special cases. 
\end{abstract}
%
%
\newpage
\noindent
\tableofcontents

\newpage
\section{Introduction}
The first and most prominent example of 
a discrete integrable system is the Toda lattice 
discovered half a century ago~\cite{Toda1,Toda2}. 
%
The 
Newtonian equations of motion 
of 
the Toda lattice are given by 
\begin{equation}
x_{n,tt} =  \mathrm{e}^{x_{n+1}-x_{n}} - \mathrm{e}^{x_{n}-x_{n-1}},
\;\;\, 
n \in \mathbb{Z}.  
\label{Toda-Newton}
\end{equation}
A simple but remarkable change of variables 
\mbox{$u_n := \frac{1}{2} \mathrm{e}^{\frac{1}{2}\left( x_{n}-x_{n-1} \right) }$}, 
\mbox{$\, w_n := \frac{1}{2} x_{n,t}$}, 
called the Flaschka--Manakov 
variables~\cite{Flaschka1,Flaschka2,Manakov74}, 
recast the Toda lattice (\ref{Toda-Newton}) 
in 
a more convenient form: 
\begin{subequations}
\label{Toda-F}
\begin{align}
& u_{n,t} = u_n (w_{n} -w_{n-1}), 
\\
& w_{n,t} = 2 \left( u_{n+1}^{2} - u_{n}^2 \right). 
\label{Toda-F2}
\end{align}
\end{subequations}
Indeed, 
the Toda lattice 
in 
Flaschka--Manakov 
variables (\ref{Toda-F}) 
is known
to have much richer 
structure than 
the original 
Newtonian equations of motion 
(\ref{Toda-Newton})~\cite{Suris03}. 
The complete integrability of the Toda lattice~\cite{Flaschka1,Flaschka2,Manakov74,Henon74} 
is based 
on the fact that (\ref{Toda-F}) is equivalent to the 
compatibility condition for the overdetermined linear system, 
generally 
called the Lax pair~\cite{Lax}: 
\begin{subequations}
\label{Toda-LM}
\begin{align}
& 
u_n \psi_{n-1} + w_n \psi_{n} + u_{n+1} \psi_{n+1} 
= \lambda \psi_{n}, 
\label{Toda-L}
\\[0mm]
& \psi_{n,t} = u_{n+1} \psi_{n+1} - u_n \psi_{n-1}, 
\label{Toda-M}
\end{align}
\end{subequations}
where $\lambda$ is 
a constant spectral parameter. 
In mathematical terms, 
the 
spatial part of the Lax pair (\ref{Toda-L}) is 
the eigenvalue problem for 
a symmetric tridiagonal matrix (Jacobi operator), 
so the Toda lattice defines an isospectral deformation of the Jacobi operator. 
In fact, (\ref{Toda-F}) 
is 
the first nontrivial 
flow of an infinite hierarchy of isospectral 
flows associated with the spectral problem (\ref{Toda-L}). 

In this paper, 
we consider an interesting extension of 
the Lax pair (\ref{Toda-LM}) 
to propose 
a 
nontrivial 
generalization of the Toda lattice hierarchy. 
This generalization 
provides 
a discrete analog of the generalization of 
the KdV hierarchy 
to 
a long wave--short wave interaction
hierarchy, called the Yajima--Oikawa hierarchy~\cite{YO76,Cheng92}, 
in the continuous case. 
A space discretization of 
the Yajima--Oikawa system was 
already proposed in the recent paper~\cite{Maruno16} 
(also see~\cite{Yu15})
and 
its Lax pair as well as 
the next higher 
symmetry 
was presented 
in~\cite{Tsuchida18-1}. 
The first flow of 
the generalized Toda hierarchy 
proposed in this paper 
provides, 
in a special case, 
a new 
integrable discretization of the 
Yajima--Oikawa 
system, 
which 
is essentially 
different from 
the discrete Yajima--Oikawa 
system studied in~\cite{Maruno16,Tsuchida18-1} 
and 
has its own advantages; 
in particular, 
the 
discrete 
Yajima--Oikawa 
system 
in this paper
possesses 
not only a Lax pair
and an infinite number of conservation laws 
but also a simple Hamiltonian structure, 
so the 
higher flows of the hierarchy 
can easily be constructed. 

This paper is organized as follows. 
In section 2, we present an extension of the 
spectral problem (\ref{Toda-L}) to a 
two-component spectral problem for $\psi_n$ and $\phi_n$, 
which can 
also 
be 
rewritten as a nonlocal spectral problem for the single component 
$\psi_n$. 
Then, we associate two isospectral 
time-evolutionary systems 
for $\psi_n$ and $\phi_n$, 
and obtain from the compatibility conditions the first 
two isospectral flows of 
the generalized 
Toda 
hierarchy. 
We also show that the first flow 
can be reduced 
to the Yajima--Oikawa 
system in a suitable continuous limit. 
In section 3, we 
prove that 
in 
some 
special (or limiting) cases, 
the first and second flows of 
the generalized Toda hierarchy 
can be reduced to 
the 
elementary flows of 
two discrete 
nonlinear Schr\"odinger hierarchies, 
the Ablowitz--Ladik 
hierarchy~\cite{AL1} 
and the Konopelchenko--Chudnovsky hierarchy~\cite{Kono82,Chud1,Chud2}, 
or 
linear 
equations. 
In section 4, we 
demonstrate 
that 
the generalized Toda hierarchy 
possesses 
two infinite sets of 
conservation laws 
and, in a special case, a simple Hamiltonian structure; 
explicit expressions for 
the higher flows 
of 
the hierarchy 
can be constructed in a recursive manner. 
Section 5 
is devoted to concluding remarks. 

\section{The generalized Toda hierarchy}
\label{sect2}

\subsection{Two-component spectral problem}

As a generalization of 
the 
eigenvalue problem for 
the 
Jacobi operator 
in (\ref{Toda-L}), 
we consider 
the following two-component spectral problem 
for $\psi_n$ and $\phi_n$: 
\begin{subnumcases}{\label{gJacobi}}
\alpha u_n^\gamma \psi_{n-1} + \beta u_{n+1}^\delta \psi_{n+1} 
	+ w_n \psi_{n} 
	+ a_n \left( \gamma \phi_n + \delta \phi_{n+1} \right) = \lambda \psi_{n}, 
\label{gToda-L1} \\[2pt]
\phi_{n+1} - \phi_n = b_n \psi_n. 
\label{gToda-L2} 
\end{subnumcases}
%
Here, $\lambda$ is 
a constant spectral parameter; 
$\alpha$, $\beta$, $\gamma$ and $\delta$ are 
arbitrary 
scalar constants
except that they 
should 
satisfy the 
conditions 
\mbox{$(\alpha \gamma, \beta \delta) \neq (0,0)$} 
and 
\mbox{$\gamma + \delta \neq 0$}. 
When the additional dependent variables $a_n$ and $b_n$ vanish, 
we recover 
the original eigenvalue problem (\ref{Toda-L}) 
by setting \mbox{$\alpha=\beta=\gamma=\delta=1$}.  
We consider the general case where 
$u_n$, 
$w_n$ 
and $\psi_n$ 
are scalar-valued functions 
and 
$a_n$, 
$b_n$ and $\phi_n$ 
are 
vector-valued functions; 
that is, 
$a_n$ is a row vector and 
$b_n$ and $\phi_n$  
are column vectors.
Note that using (\ref{gToda-L2}), we can express $\phi_n$ 
as a nonlocal function of 
$\psi_n$, 
so we can rewrite (\ref{gToda-L1}) 
as a nonlocal 
spectral problem for the single component $\psi_n$, 
{\it e.g.}, 
\begin{equation}
\alpha u_n^\gamma \psi_{n-1} + \beta u_{n+1}^\delta \psi_{n+1} 
	+\left( w_n + \delta \hspace{1pt} a_n b_n \right) \psi_{n} 
	+ \left( \gamma + \delta \right) a_n \sum_{j=-\infty}^{n-1} b_j \psi_j = \lambda \psi_{n},
\nonumber 
\end{equation}
or 
\begin{equation}
\alpha u_n^\gamma \psi_{n-1} + \beta u_{n+1}^\delta \psi_{n+1} 
	+ w_n \psi_{n} 
	+ a_n \left( \gamma \sum_{j=-\infty}^{n-1} b_j \psi_j 
	- \delta \sum_{j=n+1}^{\infty} b_j \psi_j  \right) = \lambda \psi_{n}. 
\nonumber 
\end{equation}
However, such nonlocal forms 
of the spectral problem are 
not 
convenient for 
later 
computations, 
so 
in this paper 
we 
focus on 
the 
original 
two-component form 
(\ref{gJacobi}). 

\subsection{Isospectral 
flows}
\label{subsection2.2}

As 
a trivial 
isospectral 
deformation 
of the two-component spectral problem 
(\ref{gJacobi}), 
we can consider the following:  
\begin{subnumcases}{\label{gJacobi_time0}}
\psi_{n,t_0} = c \psi_{n}, 
\label{psi-time0}
\\[2pt]
\phi_{n,t_0} = d \phi_n. 
\label{phi-time0}
\end{subnumcases}
Here, $c$ and $d$ are arbitrary scalar constants. 
The compatibility conditions for the 
overdetermined 
linear 
systems 
(\ref{gJacobi}) and (\ref{gJacobi_time0}) 
provide the trivial zeroth flow of 
the generalized Toda hierarchy: 
\begin{equation} 
\label{zeroth_flow}
\left\{ 
\begin{split}
& a_{n,t_0} -\left( c-d \right) a_n =0, 
\\[2pt]
& b_{n,t_0} + \left( c-d \right) b_n =0, 
\\[2pt]
& u_{n,t_0} =0, 
\\[2pt]
& w_{n,t_0} =0.
\end{split} 
\right. 
\end{equation}

As 
a nontrivial 
isospectral 
deformation 
of the two-component spectral problem 
(\ref{gJacobi}), 
we first consider the following:  
\begin{subnumcases}{\label{gJacobi_time1}}
\psi_{n,t_1} =  \alpha u_n^\gamma \psi_{n-1} + \beta u_{n+1}^\delta \psi_{n+1} 
	+ w_n \psi_{n}, 
\label{psi-time1}
\\[2pt]
\phi_{n,t_1} = -\alpha u_n^\gamma b_n \psi_{n-1}  + \beta u_{n}^\delta b_{n-1} \psi_{n}. 
\label{phi-time1}
\end{subnumcases}
%
The compatibility conditions for the 
overdetermined 
linear 
systems 
(\ref{gJacobi}) and (\ref{gJacobi_time1}) 
provide the first flow of 
the generalized Toda hierarchy: 
\begin{equation} 
\label{first_flow}
\left\{ 
\begin{split}
& a_{n,t_1} - \alpha u_n^\gamma a_{n-1} - \beta u_{n+1}^\delta a_{n+1} - w_n a_n =0, 
\\[2pt]
& b_{n,t_1} + \beta u_{n}^\delta b_{n-1} + \alpha u_{n+1}^\gamma b_{n+1} +b_n w_n =0, 
\\[2pt]
& u_{n,t_1} + u_n \left( a_{n-1} b_{n-1} - a_n b_n \right) =0, 
\\[2pt]
& w_{n,t_1} + \alpha \delta \left( u_n^{\gamma} a_{n-1} b_{n} 
	- u_{n+1}^{\gamma} a_n b_{n+1} \right) 
	+ \beta \gamma \left( u_n^{\delta} a_{n} b_{n-1} 
	- u_{n+1}^{\delta} a_{n+1} b_{n} \right) =0.
\end{split} 
\right. 
\end{equation}

Next, 
as another nontrivial 
isospectral 
deformation 
of the two-component spectral problem 
(\ref{gJacobi}), 
we 
consider the following: 
\begin{subnumcases}{\label{gJacobi_time2}}
\psi_{n,t_2} =  -\alpha \gamma u_n^\gamma \psi_{n-1} + \beta \delta u_{n+1}^\delta \psi_{n+1} 
	+ \gamma \delta a_n b_n \psi_{n}, 
\label{psi-time2}
\\[2pt]
\phi_{n,t_2} = \alpha \gamma u_n^\gamma b_n \psi_{n-1} + \beta \delta 
	u_{n}^\delta b_{n-1} \psi_{n}.
\label{phi-time2}
\end{subnumcases}
The compatibility conditions for the 
overdetermined 
linear 
systems 
(\ref{gJacobi}) and (\ref{gJacobi_time2}) 
provide the second flow of 
the generalized Toda hierarchy: 
\begin{equation} 
\label{second_flow}
\left\{ 
\begin{split}
& a_{n,t_2} + \alpha \gamma u_n^\gamma a_{n-1} - \beta \delta u_{n+1}^\delta a_{n+1} 
	- \gamma \delta a_n b_n a_n 
=0, 
\\[2pt]
& b_{n,t_2} + \beta \delta u_{n}^\delta b_{n-1} - \alpha \gamma u_{n+1}^\gamma b_{n+1} 
	+ \gamma \delta b_n a_n b_n 
=0, 
\\[2pt]
& u_{n,t_2} + u_n \left( w_{n-1} - w_n \right) 
	- \left( \gamma - \delta \right) u_n \left( a_{n-1} b_{n-1} - a_n b_n \right) =0, 
\\[2pt]
& w_{n,t_2} + \alpha \beta \left( \gamma + \delta \right)
	\left( u_{n}^{\gamma + \delta} - u_{n+1}^{\gamma + \delta} \right) 
	- \alpha \gamma \delta \left( u_n^{\gamma} a_{n-1} b_{n} 
 - u_{n+1}^{\gamma} a_n b_{n+1} \right) 
\\ & \hspace{5mm} + \beta \gamma \delta \left( u_n^{\delta} a_{n} b_{n-1} 
	- u_{n+1}^{\delta} a_{n+1} b_{n} \right) 
=0.
\end{split} 
\right. 
\end{equation}
%
Clearly, (\ref{second_flow}) 
is a generalization of 
the Toda lattice in Flaschka--Manakov 
variables (\ref{Toda-F}), 
wherein $a_n$ and $b_n$ are 
newly added 
dependent variables. 
Note that the second flow 
(\ref{second_flow}) can be simplified 
by 
a change of variables 
\mbox{$w_n - \left( \gamma - \delta \right) a_n b_n 
=: \widehat{w}_n$}. 

{\em Remark.} We call (\ref{first_flow}) and (\ref{second_flow}) 
the first flow and the second flow, respectively, just for convenience. 
In fact, one can call 
any linear combination 
of (\ref{first_flow}) and (\ref{second_flow}) 
the first (or second) flow. 
 
{\em Remark.} We required the condition 
\mbox{$(\alpha \gamma, \beta \delta) \neq (0,0)$} 
for the 
spectral problem 
(\ref{gJacobi}) 
to depend on 
the variable $u_n$. 
However, 
the 
above computation is valid 
even if 
this condition is 
removed; 
that is, 
in the special case \mbox{$(\alpha \gamma,\beta \delta) = (0,0)$}, 
we obtain the isospectral flows 
(\ref{zeroth_flow}), (\ref{first_flow}) and (\ref{second_flow})
without 
the equation of motion for $u_n$.

\subsection{Gauge transformation}
\label{subsec_gauge}

By applying the gauge transformation: 
\begin{equation}
\psi_n = \widehat{\psi}_n \prod_{j=-\infty}^n u_j^{\gamma}, \hspace{5mm} 
a_n = \widehat{a}_n \prod_{j=-\infty}^n u_j^{\gamma}, \hspace{5mm} 
b_n = \widehat{b}_n \prod_{j=-\infty}^n u_j^{-\gamma},
\label{gauge2}
\end{equation}
to the two-component spectral problem (\ref{gJacobi}), 
we 
obtain 
\begin{subnumcases}{\label{gJacobi4}}
\alpha \widehat{\psi}_{n-1} + \beta u_{n+1}^{\gamma+\delta} \widehat{\psi}_{n+1} 
	+ \left( w_n - \gamma \hspace{1pt} \widehat{a}_n \widehat{b}_n \right) \widehat{\psi}_{n} 
	+  \left( \gamma + \delta \right) \widehat{a}_n \phi_{n+1} 
	= \lambda \widehat{\psi}_{n}, 
\hspace{12mm}
\\[2pt]
\phi_{n+1} - \phi_n = \widehat{b}_n \widehat{\psi}_n. 
\end{subnumcases}
%
Thus, the condition \mbox{$\gamma + \delta \neq 0$} is indeed crucial 
for the 
Lax pair 
to be unfake; 
otherwise, 
$u_n$ 
is gauged away 
and 
the spectral problem does not 
involve the other dependent variables in a truly meaningful manner. 
By a redefinition of the dependent variables, 
\begin{equation}
k u_n^{\gamma+\delta} =: \widetilde{u}_n, \hspace{5mm} 
	 w_n - \gamma \hspace{1pt} \widehat{a}_n \widehat{b}_n =: \widetilde{w}_n, \hspace{5mm} 
	\left( \gamma + \delta \right) \widehat{a}_n =:  \widetilde{a}_n, 
\nonumber 
\end{equation}
we can fully remove the 
parameters $\gamma$ and $\delta$ from the 
spectral problem. 
Moreover, for nonzero values of $\alpha$ and $\beta$, we can 
apply a simple gauge transformation: 
\begin{equation}
\widehat{\psi}_n =: \alpha^{n} \widetilde{\psi}_n, \hspace{5mm} 
\widetilde{a}_n =: \alpha^{n} \widetilde{\widetilde{a}}_n, \hspace{5mm} 
\widehat{b}_n =: \alpha^{-n} \widetilde{b}_n; 
\nonumber
\end{equation}
then, 
by choosing 
the constant $k$ 
as \mbox{$k=\alpha \beta$}, 
we can 
remove the remaining parameters $\alpha$ and $\beta$ 
from the spectral problem, 
which results in the 
representative case 
\mbox{$\alpha=\beta=1$}, 
\mbox{$\gamma=0$} 
and \mbox{$\delta=1$} 
in 
(\ref{gJacobi}). 
Alternatively, we can consider the symmetric case 
\mbox{$\alpha=\beta=\gamma=\delta=1$} 
as the representative of the spectral problem (\ref{gJacobi}). 

However, 
in the following, we mainly 
consider the original general form 
(\ref{gJacobi}), 
because it encompasses some 
non-generic (say, \mbox{$\alpha=0$}) 
or limiting (say, \mbox{$\gamma + \delta \to 0$}) cases.

\subsection{Symmetry properties}
\label{subsec_symmetry}

The spectral problem (\ref{gJacobi}) and the isospectral flows 
(\ref{first_flow}) and (\ref{second_flow}) have 
two 
important 
symmetry 
properties: 
%
\begin{enumerate}
\item[(i)]
The spectral problem (\ref{gJacobi}), 
the first flow (\ref{first_flow}) and the second flow (\ref{second_flow}) 
are invariant under 
the following transformation:\\
\mbox{$\alpha \leftrightarrow \beta$}, \, \mbox{$\gamma \leftrightarrow \delta$}, \, 
\mbox{$\psi_n \to \psi_{-n}$}, \, \mbox{$\phi_n \to \phi_{-n+1}$}, \, 
\mbox{$w_n \to w_{-n}$}, \,
\mbox{$u_n \to u_{-n+1}$}, \, \mbox{$a_n \to a_{-n}$}, \, \mbox{$b_n \to -b_{-n}$}, \, 
\mbox{$t_2 \to -t_2$}. 
\item[(ii)]
The 
first flow  
(\ref{first_flow}) and the second flow (\ref{second_flow}) are 
invariant under the following transformation:\\
\mbox{$\alpha \leftrightarrow \beta$}, \, \mbox{$\gamma \leftrightarrow \delta$}, \, 
\mbox{$a_n \to \pm b_n^T$} and 
\mbox{$b_n \to \mp a_n^T$} (double sign in same order; the superscript ${}^T$ denotes 
the transpose 
of a 
vector), \, 
\mbox{$t_1 \to -t_1$}.
\end{enumerate}
%

\subsection{Complex conjugation reduction}

Reductions of an integrable system 
often 
result in 
more interesting 
systems 
than the original system 
from the point of view of 
physical or mathematical 
applications. 
For the generalized Toda hierarchy, 
we can impose a complex (or Hermitian) 
conjugation reduction 
on 
the additional dependent variables $a_n$ and $b_n$, 
so the number of dependent variables can be diminished. 

We first rescale the time variable in the 
first flow (\ref{first_flow}) as 
\begin{equation}
\partial_{t_1} =: \mathrm{i} \partial_{\tau}. 
\nonumber
\end{equation}
%
Then, 
in the simple case of scalar $u_n$, 
$w_n$, $a_n$ and $b_n$
under the parametric conditions
\begin{equation}
\beta=\alpha^\ast, \hspace{5mm} \gamma=\delta \in \mathbb{R} 
, 
\nonumber
\end{equation}
we can impose the complex conjugation reduction: 
\begin{equation}
b_n = \mathrm{i} \sigma a_n^\ast, \hspace{5mm} u_n^\ast=u_n, \hspace{5mm} w_n^\ast = w_n,  
\nonumber
\end{equation}
where $\sigma$ is 
an arbitrary real constant. 
%
For simplicity, 
we set \mbox{$\alpha=\beta=\gamma=\delta=1$} 
and \mbox{$\sigma=1$}. 
Then, this reduction simplifies the 
first flow (\ref{first_flow}) to 
\begin{equation} 
\label{reduced_first_flow}
\left\{ 
\begin{split}
& \mathrm{i} a_{n,\tau} 
= u_{n+1} a_{n+1} + w_n a_n + u_n a_{n-1} 
, 
\\[2pt]
& u_{n,\tau} = u_n \left( \left| a_n \right|^2 - \left| a_{n-1} \right|^2 \right) 
, 
\\[2pt]
& w_{n,\tau} = u_{n+1} \left( a_n a_{n+1}^\ast + a_{n+1} a_{n}^\ast \right)
 - u_n \left( a_{n-1} a_{n}^\ast + a_{n} a_{n-1}^\ast \right)
,
\end{split} 
\right. 
\end{equation}
and the second flow (\ref{second_flow}) to 
\begin{equation} 
\label{reduced_second_flow}
\left\{ 
\begin{split}
& a_{n,t_2} = u_{n+1} a_{n+1} - u_n a_{n-1} + \mathrm{i} \left| a_n \right|^2 a_n, 
\\[2pt]
& u_{n,t_2} = u_n \left( w_n - w_{n-1} \right), 
\\[2pt]
& w_{n,t_2} = 2 \left( u_{n+1}^2 - u_{n}^2 \right) 
	- \mathrm{i}  u_{n+1} \left(  a_n a_{n+1}^\ast - a_{n+1} a_{n}^\ast \right) 
+ \mathrm{i} u_n \left( a_{n-1} a_{n}^\ast - a_{n} a_{n-1}^\ast 
	 \right), 
\end{split} 
\right. 
\end{equation}
respectively, 
where \mbox{$a_n \in \mathbb{C}$} and \mbox{$u_n, w_n \in \mathbb{R}$}. 
Note that (\ref{reduced_second_flow}) 
reduces to the Toda lattice 
in 
Flaschka--Manakov 
variables (\ref{Toda-F}) by setting \mbox{$a_n
=
0$}. 

In the 
general case of 
scalar $u_n$ and 
$w_n$, 
row-vector $a_n$ and column-vector $b_n$, 
the generalized Toda hierarchy 
under the parametric conditions
\begin{equation}
\beta=\alpha^\ast, \hspace{5mm} \gamma=\delta \in \mathbb{R}, 
\nonumber
\end{equation}
admits 
the Hermitian conjugation reduction: 
\begin{equation}
b_n = \mathrm{i} \Sigma a_n^\dagger, \hspace{5mm} u_n^\ast=u_n, \hspace{5mm} w_n^\ast = w_n,  
\nonumber
\end{equation}
where $\Sigma$ is 
an arbitrary constant Hermitian matrix. 
With the aid of linear transformations 
acting on the vector components of 
$a_n$, 
we can recast $\Sigma$ 
in the canonical 
form: \mbox{$\Sigma=\mathrm{diag}\left( 1, \ldots, 1, 0, \ldots, 0, 
-1, \ldots, -1 \right)$}. 
Moreover,
if we exclude the uninteresting case of triangular (i.e., not truly coupled) systems, 
the diagonal elements of $\Sigma$ must be either $+1$ or $-1$. 
Note that the first example of an 
integrable system 
with a cubic nonlinearity of mixed signs is the vector nonlinear Schr\"odinger equation 
with both focusing and defocusing components~\cite{YO2, Ab78, New79, Mak82}. 

In the simplest case of 
\mbox{$\alpha=\beta=\gamma=\delta=1$} 
and \mbox{$\Sigma=I$},  
this reduction simplifies the 
first flow (\ref{first_flow}) 
(with \mbox{$\partial_{t_1}=
\mathrm{i} \partial_{\tau}$}) 
to 
\begin{equation} 
\label{reduced_first_flow2}
\left\{ 
\begin{split}
& \mathrm{i} a_{n,\tau} 
= u_{n+1} a_{n+1} + w_n a_n + u_n a_{n-1} 
, 
\\[2pt]
& u_{n,\tau} = u_n \left( \sca{a_n}{a_n^\ast} 
- \sca{a_{n-1}}{a_{n-1}^\ast} \right) 
, 
\\[2pt]
& w_{n,\tau} = u_{n+1} \left( \sca{a_n}{a_{n+1}^\ast} + \sca{a_{n+1}}{a_n^\ast} \right)
 - u_n \left( \sca{a_{n-1}}{a_n^\ast} + \sca{a_n}{a_{n-1}^\ast} \right)
,
\end{split} 
\right. 
\end{equation}
and the second flow (\ref{second_flow}) to 
\begin{equation} 
\label{reduced_second_flow2}
\left\{ 
\begin{split}
& a_{n,t_2} = u_{n+1} a_{n+1} - u_n a_{n-1} + \mathrm{i} \sca{a_n}{a_n^\ast} a_n, 
\\[2pt]
& u_{n,t_2} = u_n \left( w_n - w_{n-1} \right), 
\\[2pt]
& w_{n,t_2} = 2 \left( u_{n+1}^2 - u_{n}^2 \right) 
	- \mathrm{i}  u_{n+1} \left(  \sca{a_n}{a_{n+1}^\ast} - \sca{a_{n+1}}{a_n^\ast} \right) 
+ \mathrm{i} u_n \left(  \sca{a_{n-1}}{a_n^\ast} - \sca{a_n}{a_{n-1}^\ast} \right), 
\end{split} 
\right. 
\end{equation}
respectively, 
where \mbox{$a_n \in \mathbb{C}^M$} and \mbox{$u_n, w_n \in \mathbb{R}$}. 

\subsection{Continuous limit
to the 
Yajima--Oikawa 
system}

The first flow of 
the generalized Toda hierarchy 
can be reduced 
to the long wave--short wave interaction
model, called the Yajima--Oikawa system~\cite{YO76}, 
in a suitable continuous limit. 
To see this, we first rescale the dependent variable 
$a_n$ in 
(\ref{reduced_first_flow}) 
as 
\mbox{$a_n \to \Delta^{\frac{1}{2}} a_n$}, 
where $\Delta$ is a lattice 
parameter, 
to obtain 
\begin{equation} 
\label{reduced_first_flow3}
\left\{ 
\begin{split}
& \mathrm{i} a_{n,\tau} 
= 
u_{n+1} a_{n+1} + w_n a_n + u_n a_{n-1}, 
\\[2pt]
& u_{n,\tau} = \Delta u_n \left( \left| a_n \right|^2 
	- \left| a_{n-1} \right|^2 \right), 
\\[2pt]
& w_{n,\tau} = \Delta \left[ u_{n+1} \left( a_n a_{n+1}^\ast + a_{n+1} a_{n}^\ast \right)
 - u_n \left( a_{n-1} a_{n}^\ast + a_{n} a_{n-1}^\ast \right) \right]. 
\end{split} 
\right. 
\end{equation}
Alternatively, 
(\ref{reduced_first_flow3}) can be directly 
obtained 
from (\ref{first_flow}) in the scalar case by setting 
\mbox{$\partial_{t_1} = \mathrm{i} \partial_{\tau}$}, 
\mbox{$\alpha=\beta=\gamma=\delta=1$},  
\mbox{$b_n = \mathrm{i} \Delta a_n^\ast$}, 
\mbox{$u_n^\ast=u_n$} and 
\mbox{$w_n^\ast = w_n$}. 
Then, by further setting 
\[
a_n = a(n \Delta, \tau), \hspace{5mm} u_n = \frac{1}{\Delta^2} + \frac{1}{2} u(n \Delta, \tau), 
\hspace{5mm} w_n = -\frac{2}{\Delta^2} + w(n \Delta, \tau),
\]
and taking the continuous limit \mbox{$\Delta \to 0$}, 
(\ref{reduced_first_flow3}) reduces to 
\begin{equation} 
\nonumber 
\left\{ 
\begin{split}
& \mathrm{i} a_{\tau} 
= a_{xx} + \left( u+w \right) a, 
\\[2pt]
& u_{\tau} =  w_{\tau} = 2 \left( \left| a \right|^2 \right)_x, 
\end{split} 
\right. 
\end{equation}
where \mbox{$x:= n \Delta$}. 
This is indeed the Yajima--Oikawa system~\cite{YO76} for the pair of dependent 
variables \mbox{$\left( a, u+w \right)$}, 
up to a trivial 
rescaling 
and Galilean transformation. 
The Lax-pair representation for this Yajima--Oikawa system 
can 
be obtained 
from 
(\ref{gJacobi}) and (\ref{gJacobi_time1}) 
by taking the same continuous limit and thus is given by 
\begin{equation}
\nonumber 
\left\{ 
\begin{split}
& \psi_{xx} + \left( u+w \right) \psi + 2 a \phi = \lambda \psi, 
\\[2pt]
& \phi_{x} = \mathrm{i} \hspace{1pt} a^\ast \psi,  
\nonumber 
\\[2pt]
& \mathrm{i} \psi_{\tau} =  \psi_{xx} + \left( u+w \right) \psi, 
&\label{YO-psi-time1}
\nonumber 
\\[2pt]
& \phi_{\tau} = a^\ast \psi_x - a^\ast_x \psi. 
\nonumber 
\end{split}
\right. 
\end{equation}
%

In the same manner, (\ref{reduced_first_flow2}) 
can be reduced to the vector generalization 
of the Yajima--Oikawa system studied 
in~\cite{YCMa81,Mel83,MPK83,Dub88}.

\section{Special cases: discrete nonlinear Schr\"odinger hierarchies and linearization}

\subsection{Reductions}
A salient feature of 
the generalized Toda hierarchy 
is that 
one can 
equate 
the 
original 
Flaschka--Manakov variables $u_n$ and $w_n$ 
to 
some 
functions of 
the newly introduced 
variables $a_n$ and $b_n$. 
This can be 
checked by a direct calculation 
for the simpler case 
of 
\mbox{$\gamma=0$} or \mbox{$\delta=0$}. 
In view of the symmetry property (i) 
with respect to the space reflection 
as 
described in subsection~\ref{subsec_symmetry}, 
we consider 
the case \mbox{$\gamma=0$}. 

In the case \mbox{$\gamma=0$}, 
the two-component spectral problem (\ref{gJacobi}) 
reads 
\begin{subnumcases}{\label{gJacobi-gamma0}}
\alpha \psi_{n-1} + \beta u_{n+1}^\delta \psi_{n+1} 
	+ w_n \psi_{n} 
	+ \delta a_n \phi_{n+1} = \lambda \psi_{n}, 
\label{gToda-gamma0_1} \\[2pt]
\phi_{n+1} - \phi_n = b_n \psi_n, 
\label{gToda-gamma0_2} 
\end{subnumcases}
while the first flow (\ref{first_flow}) 
and the second flow (\ref{second_flow}) 
read 
\begin{equation} 
\label{first_flow-gamma0}
\left\{ 
\begin{split}
& a_{n,t_1} - \alpha a_{n-1} - \beta u_{n+1}^\delta a_{n+1} - w_n a_n =0, 
\\[2pt]
& b_{n,t_1} + \beta u_{n}^\delta b_{n-1} + \alpha b_{n+1} +b_n w_n =0, 
\\[2pt]
& u_{n,t_1} + u_n \left( a_{n-1} b_{n-1} - a_n b_n \right) =0, 
\\[2pt]
& w_{n,t_1} + \alpha \delta \left( a_{n-1} b_{n} - a_n b_{n+1} \right) =0, 
\end{split} 
\right. 
\end{equation}
and 
\begin{equation} 
\label{second_flow-gamma0}
\left\{ 
\begin{split}
& a_{n,t_2} - \beta \delta u_{n+1}^\delta a_{n+1} =0, 
\\[2pt]
& b_{n,t_2} + \beta \delta u_{n}^\delta b_{n-1} =0, 
\\[2pt]
& u_{n,t_2} + u_n \left( w_{n-1} - w_n \right) 
	+\delta u_n \left( a_{n-1} b_{n-1} - a_n b_n \right) =0, 
\\[2pt]
& w_{n,t_2} + \alpha \beta \delta 
	\left( u_{n}^{\delta} - u_{n+1}^{\delta} \right) =0,
\end{split} 
\right. 
\end{equation}
respectively. 

A direct calculation 
shows that the following 
proposition holds true. 

\begin{proposition}
\label{prop2.1}
Both \mbox{$(\ref{first_flow-gamma0})$} and \mbox{$(\ref{second_flow-gamma0})$}, 
as well as any linear combination of them 
generated 
by 
\mbox{$\partial_T := \mu \partial_{t_1} + \nu \partial_{t_2}$}, 
admit the reduction 
that $u_n^\delta$ and $w_n$ can be expressed as power series in $\alpha$ 
of the form: 
\begin{subequations}
\label{uw_expansion}
\begin{align}
\beta u_n^\delta & = k + \delta a_{n-1} b_{n} + \alpha \frac{\delta}{k} a_{n-2} \left( I + \frac{\delta}{k} b_n a_{n-1} \right) b_{n+1} 
\nonumber \\
& \hphantom{=} \; \mbox{} 
	+\alpha^2 u_n^{(2)} +\alpha^3 u_n^{(3)} + \cdots,  
\\[3mm]
w_n &= \alpha \frac{\delta}{k} a_{n-1} b_{n+1} 
	 + \alpha^2 \frac{\delta}{k^2} a_{n-2} \left( I + \frac{\delta}{k} b_n a_{n-1} \right) 
	\left( I + \frac{\delta}{k} b_{n+1} a_{n} \right) b_{n+2} 
\nonumber \\
& \hphantom{=} \; \mbox{} +\alpha^3 w_n^{(3)} +\alpha^4 w_n^{(4)}+\cdots. 
\end{align}
\end{subequations}
Here, 
$k$ is an arbitrary nonzero 
constant and $I$ is the identity matrix of the same size as $b_n a_{n-1}$. 
\end{proposition}


We conjecture that 
the power series expansion (\ref{uw_expansion}) 
is 
common to all the isospectral flows 
associated with the spectral problem (\ref{gJacobi-gamma0}) 
and all the coefficients in the power series 
%
are local functions 
of $a_n$ and $b_n$. 
 
%

Let us recall that 
the gauge transformation  (\ref{gauge2}) 
changes 
the 
spectral problem (\ref{gJacobi}) 
to the spectral problem (\ref{gJacobi4}). 
Thus, by comparing 
(\ref{gJacobi-gamma0}) with (\ref{gJacobi4}), we 
find that the replacement: 
\begin{equation}
\delta \to \gamma + \delta,
\hspace{5mm} a_n \to 
a_n \prod_{j=-\infty}^n u_j^{-\gamma}, \hspace{5mm} 
b_n \to 
b_n \prod_{j=-\infty}^n u_j^{\gamma}, \hspace{5mm} 
w_n \to w_n - \gamma \hspace{1pt} a_n b_n, 
\nonumber
\end{equation}
converts 
%
the reduction 
(\ref{uw_expansion})
in the 
special case 
\mbox{$\gamma=0$} 
to the reduction 
in the general case \mbox{$\gamma \neq 0$}. 
%
\begin{proposition}
\label{prop2.2}
The first flow \mbox{$(\ref{first_flow})$} and the second flow \mbox{$(\ref{second_flow})$}, 
as well as any linear combination of them 
generated 
by 
\mbox{$\partial_T := \mu \partial_{t_1} + \nu \partial_{t_2}$}, 
admit the reduction 
that $u_n$
and $w_n$ 
are determined 
implicitly by 
power series in $\alpha$ 
of the form: 
\begin{align}
\beta u_n^{\gamma+\delta} & = k + \left( \gamma + \delta \right) u_n^\gamma a_{n-1} b_{n} 
	+ \alpha \frac{\gamma+ \delta}{k} u_{n-1}^\gamma u_{n}^\gamma u_{n+1}^\gamma 
	a_{n-2} \left( I + \frac{\gamma + \delta}{k} u_{n}^\gamma b_n a_{n-1} \right) b_{n+1} 
\nonumber \\
& \hphantom{=} \; \mbox{} 
	+\alpha^2 u_n^{(2)} +\alpha^3 u_n^{(3)} + \cdots,  
\nonumber 
\\[3mm]
w_n &= \gamma \hspace{1pt} a_n b_n + \alpha \frac{\gamma+\delta}{k} u_{n}^\gamma u_{n+1}^\gamma a_{n-1} b_{n+1} 
\nonumber \\
& \hphantom{=} \; \mbox{} 
	 + \alpha^2 \frac{\gamma+\delta}{k^2} u_{n-1}^\gamma u_{n}^\gamma u_{n+1}^\gamma u_{n+2}^\gamma 
	a_{n-2} \left( I + \frac{\gamma+\delta}{k} u_{n}^\gamma b_n a_{n-1} \right) 
	\left( I + \frac{\gamma+\delta}{k} u_{n+1}^\gamma b_{n+1} a_{n} \right) b_{n+2} 
\nonumber \\
& \hphantom{=} \; \mbox{} +\alpha^3 w_n^{(3)} +\alpha^4 w_n^{(4)}+\cdots. 
\nonumber 
\end{align}
Here, 
$k$ is an arbitrary nonzero constant and $I$ is the identity matrix of the same size as $b_n a_{n-1}$. 
\end{proposition}

With the aid of the symmetry property (i) or (ii) 
described 
in subsection~\ref{subsec_symmetry}, 
we 
also obtain a similar reduction that 
expresses  $u_n$
and $w_n$ 
as power series in $\beta$ 
whose coefficients are 
functions 
of $a_n$ and $b_n$.

\subsection{Ablowitz--Ladik 
hierarchy
}

In the non-generic case of 
\mbox{$\alpha=\gamma=0$} or \mbox{$\beta=\delta=0$}, 
the form of the first flow (\ref{first_flow}) 
(or 
the second flow (\ref{second_flow})) 
implies 
that 
we can set \mbox{$w_n=0$} 
in the two-component spectral problem (\ref{gJacobi}) (cf.~Proposition~\ref{prop2.1}). 
In view of 
the symmetry property (i) 
described in subsection~\ref{subsec_symmetry}, 
we 
consider only the case \mbox{$\alpha=\gamma=0$}. 
Then, by 
setting \mbox{$\alpha=\gamma=0$} and \mbox{$w_n=0$}, 
the first flow (\ref{first_flow}) reduces to 
%
\begin{equation} 
\label{first_flow_reduced}
\left\{ 
\begin{split}
& a_{n,t_1} - \beta u_{n+1}^\delta a_{n+1} =0, 
\\[2pt]
& b_{n,t_1} + \beta u_{n}^\delta b_{n-1} =0, 
\\[2pt]
& u_{n,t_1} + u_n \left( a_{n-1} b_{n-1} - a_n b_n \right) =0,
\end{split} 
\right. 
\end{equation}
and the second flow (\ref{second_flow}) 
reduces to 
\begin{equation} 
\label{second_flow_reduced}
\left\{ 
\begin{split}
& a_{n,t_2} - \beta \delta u_{n+1}^\delta a_{n+1} 
=0, 
\\[2pt]
& b_{n,t_2} + \beta \delta u_{n}^\delta b_{n-1} 
=0, 
\\[2pt]
& u_{n,t_2} +\delta u_n \left( a_{n-1} b_{n-1} - a_n b_n \right) =0,
\end{split} 
\right. 
\end{equation}
respectively. 
We 
recall the condition \mbox{$\beta \delta \neq 0$} 
for the spectral problem (\ref{gJacobi}) 
with \mbox{$\alpha=\gamma=0$} and \mbox{$w_n=0$} 
(and consequently (\ref{first_flow_reduced}) and (\ref{second_flow_reduced}))
to depend on $u_n$. 
Then, 
(\ref{second_flow_reduced}) 
is equivalent to (\ref{first_flow_reduced})
up to a rescaling of the time variable, so 
we 
consider only (\ref{first_flow_reduced}). 
Note that (\ref{first_flow_reduced}) with \mbox{$\beta=-1$} and \mbox{$\delta=1$} 
coincides with (4.19) or (4.21) in \cite{Li2016}. 
Because (\ref{first_flow_reduced}) 
implies the relation:
\[
\left( \beta u_n^\delta - \delta a_{n-1} b_n \right)_{t_1} =0,
\]
we can set 
\begin{equation}
\beta u_n^\delta = k_n + \delta a_{n-1} b_n, 
\nonumber 
\end{equation}
where $k_n$ is independent of time $t_1$. 
Thus, the 
three-component system 
(\ref{first_flow_reduced}) is simplified to 
the two-component system: 
\begin{equation} 
\nonumber 
\left\{ 
\begin{split}
& a_{n-1,t_1} - \left( k_n + \delta a_{n-1} b_n \right) a_{n} =0, 
\\[2pt]
& b_{n,t_1} + \left( k_n + \delta a_{n-1} b_n \right) b_{n-1} =0. 
\end{split} 
\right. 
\end{equation}
By a rescaling of $a_{n}$ and $b_{n}$, 
we can normalize $\delta$ to $1$. 
Moreover, for nonzero values of $k_n$, 
$k_n$ can be 
fixed at $1$ 
by 
the 
simple transformation 
\begin{equation} 
a_{n-1} =  \left( \prod_{j
}^{n-1} \frac{1}{k_j} \right)
	\vt{a}_{n}, 
\hspace{5mm} 
 b_{n} = \left( \prod_{j
}^{n} k_j \right) \vt{b}_{n}.  
\nonumber  
\end{equation}
Here, $\vt{a}_n$ is a row vector and $\vt{b}_n$ is a column vector.
Thus, we finally obtain (the straightforward vector generalization~\cite{GI82} of) 
an elementary 
flow of 
the Ablowitz--Ladik hierarchy~\cite{AL1
}:
\begin{equation} 
\label{ab_relation2}
\left\{ 
\begin{split}
& \vt{a}_{n,t_1} - \left( 1 + \vt{a}_{n}\vt{b}_n \right) \vt{a}_{n+1}=0, 
\\[2pt]
& \vt{b}_{n,t_1} + \left( 1 + \vt{a}_{n}\vt{b}_n \right) \vt{b}_{n-1}=0. 
\end{split} 
\right. 
\end{equation} 

The derivation given above 
implies 
that the Lax-pair representation for 
the Ablowitz--Ladik 
flow (\ref{ab_relation2}) is given by 
the two-component spectral problem 
\begin{subnumcases}{\label{2com-AL}}
 \left( 1 + \vt{a}_{n+1}\vt{b}_{n+1} \right) \psi_{n+1} 
	+ \vt{a}_{n+1}  \phi_{n+1} = \lambda \psi_{n}, 
\label{AL-L1} \\[2pt]
\phi_{n+1} - \phi_n = \vt{b}_{n} \psi_n,
\label{AL-L2} 
\end{subnumcases}
and the isospectral 
time-evolutionary system
\begin{equation}
\nonumber 
\left\{ 
\begin{split}
& \psi_{n,t_1} = \left( 1 + \vt{a}_{n+1}\vt{b}_{n+1} \right) \psi_{n+1},  
\\[2pt]
& \phi_{n,t_1} =  \left( 1 + \vt{a}_{n}\vt{b}_{n} \right) \vt{b}_{n-1} \psi_{n},  
\end{split}
\right. 
\end{equation}
where $\psi_n$ 
is a scalar 
and $\phi_n$ is a column vector. 

To obtain the other elementary 
flow of  
the Ablowitz--Ladik hierarchy, 
we set \mbox{$\gamma=0$} 
and consider a suitable linear combination of the first flow (\ref{first_flow-gamma0})
and the second flow (\ref{second_flow-gamma0}) 
in the case 
\mbox{$\alpha \sim 0$}. 
Indeed, by setting 
\begin{equation}
\partial_T := \frac{1}{\alpha \delta} \left( \partial_{t_2} - \delta \partial_{t_1} \right), 
\nonumber
\end{equation}
%
and imposing 
the reduction stated in 
Proposition~\ref{prop2.1},
\begin{align}
& \beta u_n^\delta = k + \delta a_{n-1} b_{n} + O(\alpha), 
\nonumber 
\\[1mm]
& \frac{1}{\alpha} w_n = \frac{\delta}{k} a_{n-1} b_{n+1} + O( \alpha ), 
\nonumber 
\end{align}
we have 
\begin{equation} 
\nonumber
\left\{ 
\begin{split}
& a_{n, T} + a_{n-1} + \left( \frac{\delta}{k} a_{n-1} b_{n+1} + O( \alpha ) \right) a_{n} =0, 
\\[2pt]
& b_{n,T} - b_{n+1} - b_n \left( \frac{\delta}{k} a_{n-1} b_{n+1} + O( \alpha ) \right) =0. 
\end{split} 
\right. 
\end{equation}
Thus, 
by taking the limit \mbox{$\alpha \to 0$}, 
setting 
\mbox{$k=\delta=1$} and writing 
\mbox{$a_{n-1} = \vt{a}_{n}$} and \mbox{$b_{n} = \vt{b}_{n}$}, 
we obtain 
(the 
vector generalization~\cite{GI82} of) 
the other elementary 
flow of  
the Ablowitz--Ladik hierarchy~\cite{AL1
}: 
%
\begin{equation} 
\label{ab_relation4}
\left\{ 
\begin{split}
& \vt{a}_{n,T} + \vt{a}_{n-1} + \vt{a}_{n-1}\vt{b}_n \vt{a}_{n}=0, 
\\[2pt]
& \vt{b}_{n,T} - \vt{b}_{n+1} - \vt{b}_n \vt{a}_{n} \vt{b}_{n+1}=0. 
\end{split} 
\right. 
\end{equation} 
%
The Lax-pair representation for (\ref{ab_relation4}) is given by the two-component 
spectral problem 
(\ref{2com-AL}) and the isospectral time-evolutionary system 
\begin{equation}
\nonumber 
\left\{ 
\begin{split}
& \psi_{n,T} = -\vt{a}_{n}\vt{b}_{n+1} \psi_{n} - \psi_{n-1},  
\\[2pt]
& \phi_{n,T} = \vt{b}_{n} \psi_{n-1}.
\end{split}
\right. 
\end{equation}

%
%

\subsection{Konopelchenko--Chudnovsky 
hierarchy}

Let us first recall the remark at the end of subsection~\ref{subsection2.2}. 
%
%
%
In 
the 
special case \mbox{$\alpha=\delta=0$}, 
the first flow (\ref{first_flow}) 
(without the equation of motion for $u_n$) 
reads~\cite{Zhang91,Li2016}
\begin{equation} 
\label{dNLS_flow}
\left\{ 
\begin{split}
& a_{n,t_1} - \beta a_{n+1} - w_n a_n =0, 
\\[2pt]
& b_{n,t_1} + \beta b_{n-1} +b_n w_n =0, 
\\[2pt]
& w_{n,t_1} + \beta \gamma \left(  a_{n} b_{n-1} - a_{n+1} b_{n} \right) =0, 
\end{split} 
\right. 
\end{equation}
which implies the relation: 
\begin{equation}
 \left( w_n - \gamma a_n b_n \right)_{t_1} =0. 
\nonumber
\end{equation}
Thus, we can set 
\begin{equation}
w_n = \mu_n + \gamma a_n b_n, 
\nonumber 
\end{equation}
where $\mu_n$ 
is an arbitrary function of $n$, 
independent of time $t_1$. 
Then, the 
three-component system (\ref{dNLS_flow}) is simplified to the two-component system:
\begin{equation} 
\label{dNLS_flow2}
\left\{ 
\begin{split}
& a_{n,t_1} - \beta a_{n+1} - \left( \mu_n + \gamma a_n b_n 
\right) a_n =0, 
\\[2pt]
& b_{n,t_1} + \beta b_{n-1} +b_n \left( \mu_n + \gamma a_n b_n 
\right) =0.
\end{split} 
\right. 
\end{equation}
If we 
interpret $t_1$ as the continuous spatial variable $x$, 
(\ref{dNLS_flow2}) 
can be identified with 
an elementary auto-B\"acklund
transformation for the continuous nonlinear Schr\"odinger hierarchy 
studied by Konopelchenko~\cite{Kono82} and 
D.~V.\ and G.~V.~Chudnovsky~\cite{Chud1,Chud2}; 
in this context, the $n$-dependence of $\mu_n$ is essential and 
not removable by any simple transformation. 
In the simplest case 
\mbox{$\mu_n=0$}, 
(\ref{dNLS_flow2}) 
provides 
an elementary flow~\cite{GI82} of 
an integrable discrete 
nonlinear Schr\"odinger hierarchy, which we call the Konopelchenko--Chudnovsky hierarchy~\cite{Kono82,Chud1,Chud2}. 

%
To obtain the other elementary flow~\cite{GI82,Tu} of the Konopelchenko--Chudnovsky hierarchy, 
we consider the second flow (\ref{second_flow}) 
in the case \mbox{$\delta=0$}, 
\mbox{$\alpha \sim 0$}. 
The second flow (\ref{second_flow}) 
with \mbox{$\delta=0$} 
reads 
\begin{equation} 
\label{second_flow_delta0}
\left\{ 
\begin{split}
& a_{n,t_2} + \alpha \gamma u_n^\gamma a_{n-1} =0, 
\\[2pt]
& b_{n,t_2} - \alpha \gamma u_{n+1}^\gamma b_{n+1} =0, 
\\[2pt]
& u_{n,t_2} + u_n \left( w_{n-1} - w_n \right) - \gamma  u_n \left( a_{n-1} b_{n-1} - a_n b_n \right) =0, 
\\[2pt]
& w_{n,t_2} + \alpha \beta \gamma \left( u_{n}^{\gamma} - u_{n+1}^{\gamma} \right) =0.
\end{split} 
\right. 
\end{equation}
We rescale the time derivative as 
\[
\partial_{t_{-1}} := \frac{1}{\alpha \gamma} \partial_{t_2},
\]
and consider 
the reduction guaranteed by Proposition~\ref{prop2.2}: 
\begin{align}
& u_n^{\gamma} = \frac{k}{\beta - \gamma a_{n-1} b_{n}} + O( \alpha),  
\nonumber 
\\[1mm]
& w_n = \gamma a_n b_n + O(\alpha). 
\nonumber 
\end{align}
%
Thus, in the limit \mbox{$\alpha \to 0$}, 
(\ref{second_flow_delta0}) under this reduction 
reduces to the other elementary flow~\cite{GI82,Tu}
of the Konopelchenko--Chudnovsky hierarchy: 
\begin{equation} 
\nonumber
\left\{ 
\begin{split}
& a_{n,t_{-1}} + \frac{k}{\beta - \gamma a_{n-1} b_{n}} a_{n-1} =0, 
\\[2pt]
& b_{n,t_{-1}} - \frac{k}{\beta - \gamma a_{n} b_{n+1}} b_{n+1} =0. 
\end{split} 
\right. 
\end{equation}


\subsection{Exceptional 
case:\ linearization}


For the 
spectral problem 
(\ref{gJacobi}) 
to involve 
the dependent variables 
in a meaningful manner, 
we assumed  
the condition 
\mbox{$\gamma + \delta \neq 0$} 
(see subsection~\ref{subsec_gauge}). 
Thus, the case \mbox{$\gamma + \delta = 0$} is exceptional and excluded 
from our consideration. 
In fact, 
the first flow (\ref{first_flow}) and the second flow (\ref{second_flow}) of 
the generalized Toda hierarchy 
in this exceptional case 
turn out to be 
linearizable. 

The first flow (\ref{first_flow}) 
with \mbox{$ \delta =-\gamma$} is 
\begin{equation} 
\label{first_flow_linearize}
\left\{ 
\begin{split}
& a_{n,t_1} - \alpha u_n^\gamma a_{n-1} - \beta u_{n+1}^{-\gamma} a_{n+1} - w_n a_n =0, 
\\[2pt]
& b_{n,t_1} + \beta u_{n}^{-\gamma} b_{n-1} + \alpha u_{n+1}^\gamma b_{n+1} +b_n w_n =0, 
\\[2pt]
& u_{n,t_1} + u_n \left( a_{n-1} b_{n-1} - a_n b_n \right) =0, 
\\[2pt]
& w_{n,t_1} - \alpha \gamma \left( u_n^{\gamma} a_{n-1} b_{n} 
	- u_{n+1}^{\gamma} a_n b_{n+1} \right) 
	+ \beta \gamma \left( u_n^{-\gamma} a_{n} b_{n-1} 
	- u_{n+1}^{-\gamma} a_{n+1} b_{n} \right) =0.
\end{split} 
\right. 
\end{equation}
From (\ref{first_flow_linearize}), 
we obtain the relations: 
\begin{equation}
 \left\{
\exp \left[ \int \left( \lim_{n \to - \infty}  a_n b_n \right) \mathrm{d} t_1 \right] 
 \prod_{j=-\infty}^{n} u_j \right\}_{t_1} = a_n b_n \left\{ \exp \left[ \int \left( \lim_{n \to - \infty}  a_n b_n \right) \mathrm{d} t_1 \right] 
 \prod_{j=-\infty}^{n} u_j \right\}, 
\nonumber
\end{equation}
and
\begin{equation}
 \left( w_n - \gamma a_n b_n \right)_{t_1} =0. 
\nonumber
\end{equation}
Thus, 
the new dependent variables defined as 
\begin{align}
& \mathcal{A}_n := 
 \left\{ \exp \left[ \int \left( \lim_{n \to - \infty}  a_n b_n \right) \mathrm{d} t_1 \right] 
 \prod_{j=-\infty}^{n} u_j \right\}^{-\gamma} a_n, 
\nonumber \\[2mm]
& \mathcal{B}_n := 
b_n \left\{ \exp \left[ \int \left( \lim_{n \to - \infty}  a_n b_n \right) \mathrm{d} t_1 \right] 
 \prod_{j=-\infty}^{n} u_j \right\}^{\gamma}, 
\nonumber 
\end{align}
satisfy the linear equations with $n$-dependent 
(but $t_1$-independent) coefficients: 
\begin{equation} 
\nonumber
\left\{ 
\begin{split}
& \mathcal{A}_{n,t_1} 
	- \alpha \mathcal{A}_{n-1} - \beta \mathcal{A}_{n+1} 
	- \left( w_n - \gamma a_n b_n \right)  \mathcal{A}_{n}  =0, 
\\[2pt]
& \mathcal{B}_{n,t_1} 
	+ \beta \mathcal{B}_{n-1} + \alpha \mathcal{B}_{n+1} 
	+ \mathcal{B}_n \left( w_n - \gamma a_n b_n \right) =0. 
\end{split} 
\right. 
\end{equation}

The second flow (\ref{second_flow}) with \mbox{$ \delta =-\gamma$}: 
\begin{equation} 
\nonumber
\left\{ 
\begin{split}
& a_{n,t_2} + \alpha \gamma u_n^\gamma a_{n-1} + \beta \gamma u_{n+1}^{-\gamma} a_{n+1} 
	+ \gamma^2 a_n b_n a_n 
=0, 
\\[2pt]
& b_{n,t_2} - \beta \gamma u_{n}^{-\gamma} b_{n-1} - \alpha \gamma u_{n+1}^\gamma b_{n+1} 
	- \gamma^2 b_n a_n b_n 
=0, 
\\[2pt]
& u_{n,t_2} + u_n \left( w_{n-1} - w_n \right) 
	- 2 \gamma u_n \left( a_{n-1} b_{n-1} - a_n b_n \right) =0, 
\\[2pt]
& w_{n,t_2} +\alpha \gamma^2 \left( u_n^{\gamma} a_{n-1} b_{n} 
 - u_{n+1}^{\gamma} a_n b_{n+1} \right) 
- \beta \gamma^2 \left( u_n^{-\gamma} a_{n} b_{n-1} 
	- u_{n+1}^{-\gamma} a_{n+1} b_{n} \right) 
=0, 
\end{split} 
\right. 
\end{equation}
can be linearized in a similar manner. 

\section{Hamiltonian structure and Conservation laws}

In the special case \mbox{$\gamma=\delta$}, 
the generalized Toda hierarchy 
possesses a local 
Hamiltonian structure; 
the trivial zeroth flow 
(\ref{zeroth_flow}), 
the 
first flow (\ref{first_flow}) and the second flow (\ref{second_flow}) 
with \mbox{$\gamma=\delta$} 
can be written in the 
Hamiltonian form: 
%
\begin{align} 
\nonumber
a_{n,t_l}^{(j)} = \left\{ a_n^{(j)}, H_l \right\},  \hspace{4mm}
b_{n,t_l}^{(j)} = \left\{ b_n^{(j)}, H_l \right\},  \hspace{4mm}
u_{n,t_l} = \left\{ u_n, H_l \right\}, \hspace{4mm}
w_{n,t_l} = \left\{ w_n, H_l \right\}, 
\\ 
\nonumber 
l= 0, 1, 2, 
\end{align}
%
where $a_n^{(j)}$ is the $j$th component of the row vector $a_n$ 
and 
$b_n^{(j)}$ is the $j$th component of the column vector $b_n$. 
The set of nonvanishing 
Poisson brackets and Hamiltonians are 
given by 
\begin{equation} 
\label{gPB}
\left\{ a_n^{(j)}, b_n^{(k)} \right\}= \frac{1}{\kappa} \delta_{jk},  
\hspace{5mm}
\left\{ u_n, w_n \right\} = \frac{1}{\kappa} u_n, \hspace{4mm}
\left\{ u_n, w_{n-1} \right\} = -\frac{1}{\kappa} u_n,
\end{equation}
and 
\begin{subequations}
\label{H_012}
\begin{align}
& H_0 = \kappa \left( c-d \right) \sum_n a_n b_n,
\\[2mm]
& H_1 = \kappa \sum_n \left( w_n a_n b_n + \alpha u_{n}^\gamma a_{n-1} b_{n} + \beta u_n^\gamma a_n b_{n-1} \right),
\\[2mm]
& H_2 = \kappa \sum_n \left[ \frac{1}{2} w_n^2 + \frac{1}{2} \gamma^2 \left( a_n b_n \right)^2 
	- \alpha \gamma u_{n}^\gamma a_{n-1} b_{n} + \beta \gamma u_n^\gamma a_n b_{n-1} + \alpha \beta u_n^{2\gamma} \right] ,
\end{align}
\end{subequations}
respectively. 
Here, $\kappa$ is an arbitrary 
nonzero 
constant 
and 
$\delta_{jk}$ is 
the Kronecker delta; 
$\sum_n \log u_n$ and 
$\sum_n w_n$ 
are Casimir functions. 
Note that 
(\ref{gPB}) 
(especially with \mbox{$\kappa =4$})
is a natural extension 
of the 
canonical Poisson 
bracket 
relations 
for the Toda lattice 
written in 
Flaschka--Manakov 
variables~\cite{Flaschka1,Flaschka2,Manakov74}. 
 

We present a recursive method for 
constructing 
two infinite sets of 
conservation laws 
of 
the generalized Toda hierarchy; 
in the 
special 
case \mbox{$\gamma=\delta$}, 
the method 
allows 
us to generate 
the higher flows of the hierarchy 
with the aid of the Poisson 
bracket relations 
(\ref{gPB}). 

First, we set 
\begin{equation}
\frac{\psi_{n}}{\psi_{n+1}} =: \frac{\beta}{\lambda} u_{n+1}^\delta J_n ,
	\hspace{5mm} \frac{\phi_{n+1}}{\psi_n} =: K_n ,
\label{JK_def}
\end{equation}
and rewrite 
the two-component spectral problem (\ref{gJacobi}) 
as the relations for $J_n$ and $K_n$: 
\begin{subnumcases}{\label{YO-JK}}
J_n = 1+ \frac{1}{\lambda}\left( w_n - \gamma a_n b_n \right) J_n 
	+ \frac{\gamma + \delta}{\lambda} a_n K_n J_n + \frac{\alpha \beta}{\lambda^2} u_n^{\gamma + \delta} J_{n-1} J_{n}, \;\;\;
\label{YO-JK1}
\\[2pt]
K_n = b_n + \frac{\beta}{\lambda} u_{n}^\delta K_{n-1} J_{n-1}. 
\label{YO-JK2}
\end{subnumcases}
Here, the condition \mbox{$\beta \neq 0$} is assumed 
to derive (\ref{YO-JK1}), 
but 
this assumption is not essential 
and 
can in fact be removed. 
Using the relations (\ref{YO-JK}) recursively, 
we can express $J_n$ and $K_n$ as power series in $1/\lambda$: 
\begin{align}
J_n = \sum_{j=0}^\infty \frac{1}{\lambda^j} J_n^{(j)}, 
\hspace{5mm}
K_n =  \sum_{j=0}^\infty \frac{1}{\lambda^j} K_n^{(j)}. 
\nonumber 
\end{align}
%
More specifically, 
we have the following 
power series expansions: 
%
\begin{align}
J_n &= 1+ \frac{1}{\lambda} \left( w_n + \delta a_n b_n \right) 
+\frac{1}{\lambda^2} \left[ \left( w_n + \delta a_n b_n \right)^2 
	+ \left( \gamma +\delta \right) \beta u_n^\delta a_n b_{n-1} + \alpha \beta u_n^{\gamma + \delta} \right]
\nonumber \\[2pt]
& \hphantom{=} \; \mbox{}
+\frac{1}{\lambda^3} \left\{ \left( w_n + \delta a_n b_n \right) \left[ \left( w_n + \delta a_n b_n \right)^2 
	+ \left( \gamma +\delta \right) \beta u_n^\delta a_n b_{n-1} + \alpha \beta u_n^{\gamma + \delta} \right] 
\right. 
\nonumber \\[2pt] 
& \hphantom{=} \; \mbox{} 
+  \left( \gamma +\delta \right) \beta u_n^\delta a_n b_{n-1} \left( w_n + \delta a_n b_n \right) 
\nonumber
\\[2pt] 
& \hphantom{=} \; \mbox{} 
+ \left( \gamma +\delta \right) \beta u_n^\delta \left[ \beta u_{n-1}^\delta a_n b_{n-2} + a_n b_{n-1} 
	\left( w_{n-1} +\delta a_{n-1}b_{n-1} \right) \right] 
\nonumber \\[2pt] 
& \hphantom{=} \; \mbox{} + \left. \alpha \beta u_n^{\gamma+\delta} \left( w_{n-1} +\delta a_{n-1}b_{n-1} + w_n + \delta a_n b_n\right)
\right\}
\nonumber \\[2pt]
& \hphantom{=} \; \mbox{}+	O\left( \frac{1}{\lambda^4} \right), 
\nonumber
\\[5pt]
K_n &= b_n +\frac{1}{\lambda} \beta u_n^\delta b_{n-1} 
+ \frac{1}{\lambda^2} \beta u_n^\delta
\left[ \beta u_{n-1}^\delta b_{n-2} + b_{n-1} 
	\left( w_{n-1} +\delta a_{n-1}b_{n-1} \right)
\right]
\nonumber \\[2pt]
\nonumber 
& \hphantom{=} \; \mbox{}+	O\left( \frac{1}{\lambda^3} \right). 
\end{align}
We 
consider 
the 
identity~\cite{Wadati1977,Haberman}:
\begin{equation}
\left[ \log \left( \frac{\psi_n}{\psi_{n+1}} \right) \right]_{t_j} 
+\boldsymbol{\Delta}_n^+ \left[ \frac{\psi_{n,t_j}}{\psi_{n}} \right] =0, 
\hspace{5mm} j=0,1,2, 
\label{start_point}
\end{equation}
where 
$\boldsymbol{\Delta}_n^+$ 
is the forward difference operator 
(\mbox{$ \boldsymbol{\Delta}_n^+ f_{n} 
:=
f_{n+1}
-f_{n}$}); 
(\ref{start_point}) can be rewritten 
with the aid of (\ref{JK_def}) and (\ref{psi-time0}), (\ref{psi-time1}) or (\ref{psi-time2}) 
as 
the
conservation 
laws: 
\begin{subequations}
\label{cons_generate1}
\begin{align}
& 
\left[ \delta \log u_{n+1} + \log J_n \right]_{t_0} 
=0,
\label{generating1}
\\[2mm]
& \left[ \delta \log u_{n+1} + \log J_n \right]_{t_1} + 
\boldsymbol{\Delta}_n^+ \left[ \frac{\alpha\beta}{\lambda} u_{n}^{\gamma+\delta} J_{n-1} +\frac{\lambda}{J_n} +w_n \right] 
=0,
\label{generating2}
\\[2mm]
& \left[ \delta \log u_{n+1} + \log J_n \right]_{t_2} + 
\boldsymbol{\Delta}_n^+ \left[ -\frac{\alpha\beta\gamma}{\lambda} u_{n}^{\gamma+\delta} J_{n-1} +\frac{\delta \lambda}{J_n} 
	+ \gamma \delta a_n b_n \right] 
=0. 
\label{generating3}
\end{align}
\end{subequations}
Substituting the power series expansion 
for $J_n$ 
into (\ref{cons_generate1}) 
and equating the coefficients of different 
powers of $1/\lambda$ on the left-hand side to zero, we 
obtain an infinite set of conservation laws 
for the first three flows 
of 
the generalized Toda hierarchy. 
Note that 
\mbox{$\delta \log u_{n+1} + \log J_n $}
is 
a generating function of the 
conserved densities; 
the first four 
conserved densities obtained in this 
manner 
are 
\begin{align}
 I^{(0)}_{n} &= \log u_{n}, 
\nonumber \\[2mm]
 I^{(1)}_{n} &= 
	w_n + \delta a_n b_n, 
\nonumber \\[2mm]
 I^{(2)}_{n} &= 
	\frac{1}{2} \left( w_n + \delta a_n b_n \right)^2 + \left( \gamma + \delta \right) \beta u_n^\delta a_n b_{n-1} 
	+ \alpha \beta u_n^{\gamma + \delta}, 
\nonumber \\[2mm]
  I^{(3)}_{n} &= 
	\left[ \left( \gamma + \delta \right) \beta u_n^\delta a_n b_{n-1} + \alpha \beta u_n^{\gamma + \delta} \right] 
	\left( w_n + \delta a_n b_n 
	+ w_{n-1} + \delta a_{n-1} b_{n-1} \right) 
\nonumber \\
& \hphantom{=} \; \mbox{}+\left( \gamma + \delta \right) \beta^2 u_n^\delta u_{n-1}^\delta a_n b_{n-2}
	+ \frac{1}{3} \left( w_n + \delta a_n b_n \right)^3. 
\nonumber 
\end{align}

Second, we set 
\begin{equation}
\frac{\psi_{n}}{\psi_{n-1}} =: \frac{\alpha}{\lambda} u_{n}^\gamma \mathcal{J}_n ,
	\hspace{5mm} \frac{\phi_{n}}{\psi_n} =: \mathcal{K}_n ,
\label{calJK_def}
\end{equation}
and rewrite 
the two-component spectral problem (\ref{gJacobi}) 
as the relations for $\mathcal{J}_n$ and $\mathcal{K}_n$: 
\begin{subnumcases}{\label{YO-calJK}}
\mathcal{J}_n = 1+ \frac{1}{\lambda}\left( w_n +\delta a_n b_n \right) \mathcal{J}_n 
	+ \frac{\gamma + \delta}{\lambda} a_n \mathcal{K}_n \mathcal{J}_n 
	+ \frac{\alpha \beta}{\lambda^2} u_{n+1}^{\gamma + \delta} \mathcal{J}_{n+1} \mathcal{J}_{n}, \;\;\;
\label{YO-calJK1}
\\[2pt]
\mathcal{K}_n = -b_n + \frac{\alpha}{\lambda} u_{n+1}^\gamma \mathcal{K}_{n+1} \mathcal{J}_{n+1}. 
\label{YO-calJK2}
\end{subnumcases}
Here, the condition \mbox{$\alpha \neq 0$} is assumed to derive (\ref{YO-calJK1}), 
but this assumption is not essential and can in fact be removed. 
Using the relations (\ref{YO-calJK}) recursively, 
we can express $\mathcal{J}_n$ and $\mathcal{K}_n$ as power series in $1/\lambda$: 
\begin{align}
\mathcal{J}_n = \sum_{j=0}^\infty \frac{1}{\lambda^j} \mathcal{J}_n^{(j)}, 
\hspace{5mm}
\mathcal{K}_n =  \sum_{j=0}^\infty \frac{1}{\lambda^j} \mathcal{K}_n^{(j)}. 
\nonumber 
\end{align}
%
More specifically, 
we have the following 
power series expansions: 
\begin{align}
\mathcal{J}_n &= 1+ \frac{1}{\lambda} \left( w_n - \gamma a_n b_n \right) 
+\frac{1}{\lambda^2} \left[ \left( w_n - \gamma a_n b_n \right)^2 
	- \left( \gamma +\delta \right) \alpha u_{n+1}^\gamma a_n b_{n+1} + \alpha \beta u_{n+1}^{\gamma + \delta} \right]
\nonumber \\[2pt]
& \hphantom{=} \; \mbox{}
+\frac{1}{\lambda^3} \left\{ \left( w_n - \gamma a_n b_n \right) \left[ \left( w_n -\gamma a_n b_n \right)^2 
	- \left( \gamma +\delta \right) \alpha u_{n+1}^\gamma a_n b_{n+1} + \alpha \beta u_{n+1}^{\gamma + \delta} \right] 
\right. 
\nonumber \\[2pt] 
& \hphantom{=} \; \mbox{} 
-  \left( \gamma +\delta \right) \alpha u_{n+1}^\gamma a_n b_{n+1} \left( w_n - \gamma a_n b_n \right) 
\nonumber
\\[2pt] 
& \hphantom{=} \; \mbox{} 
- \left( \gamma +\delta \right) \alpha u_{n+1}^\gamma \left[ \alpha u_{n+2}^\gamma a_n b_{n+2} + a_n b_{n+1} 
	\left( w_{n+1} -\gamma a_{n+1}b_{n+1} \right) \right] 
\nonumber \\[2pt] 
& \hphantom{=} \; \mbox{} + \left. \alpha \beta u_{n+1}^{\gamma+\delta} \left( w_{n+1} -\gamma a_{n+1}b_{n+1} 
	+ w_n -\gamma a_n b_n\right)
\right\}
\nonumber \\[2pt]
& \hphantom{=} \; \mbox{}+	O\left( \frac{1}{\lambda^4} \right), 
\nonumber
\\[5pt]
\mathcal{K}_n &= -b_n -\frac{1}{\lambda} \alpha u_{n+1}^\gamma b_{n+1} 
- \frac{1}{\lambda^2} \alpha u_{n+1}^\gamma
\left[ \alpha u_{n+2}^\gamma b_{n+2} + b_{n+1} 
	\left( w_{n+1} -\gamma a_{n+1}b_{n+1} \right)
\right]
\nonumber \\[2pt]
\nonumber 
& \hphantom{=} \; \mbox{}+	O\left( \frac{1}{\lambda^3} \right). 
\end{align}
We 
consider 
the 
identity~\cite{Wadati1977,Haberman}:
\begin{equation}
\left[ \log \left( \frac{\psi_n}{\psi_{n-1}} \right) \right]_{t_j} 
-\boldsymbol{\Delta}_n^+ \left[ \frac{\psi_{n-1,t_j}}{\psi_{n-1}} \right] =0, 
\hspace{5mm} j=0,1,2, 
\label{start_point2}
\end{equation}
where 
$\boldsymbol{\Delta}_n^+$ 
is the forward difference operator;  
(\ref{start_point2}) can be rewritten 
with the aid of (\ref{calJK_def}) and (\ref{psi-time0}), (\ref{psi-time1}) or (\ref{psi-time2}) 
as 
the
conservation 
laws: 
\begin{subequations}
\label{cons_generate1'}
\begin{align}
& 
\left[ \gamma \log u_{n} + \log \mathcal{J}_n \right]_{t_0} 
=0,
\label{generating1'}
\\[2mm]
& \left[ \gamma \log u_{n} + \log \mathcal{J}_n \right]_{t_1} -
\boldsymbol{\Delta}_n^+ \left[ \frac{\lambda}{\mathcal{J}_{n-1}} + \frac{\alpha\beta}{\lambda} u_{n}^{\gamma+\delta} \mathcal{J}_{n} 
	+w_{n-1} \right] 
=0,
\label{generating2'}
\\[2mm]
& \left[ \gamma \log u_{n} + \log \mathcal{J}_n \right]_{t_2} - 
\boldsymbol{\Delta}_n^+ \left[ -\frac{\gamma \lambda}{\mathcal{J}_{n-1}} 
	+\frac{\alpha\beta\delta}{\lambda} u_{n}^{\gamma+\delta} \mathcal{J}_{n} 
	+ \gamma \delta a_{n-1} b_{n-1} \right] 
=0. 
\label{generating3'}
\end{align}
\end{subequations}
Substituting the power series expansion 
for $\mathcal{J}_n$ 
into (\ref{cons_generate1'}) 
and equating the coefficients of different 
powers of $1/\lambda$ on the left-hand side to zero, we 
obtain an infinite set of conservation laws 
for the first three flows 
of 
the generalized Toda hierarchy. 
Note that 
\mbox{$\gamma \log u_{n} + \log \mathcal{J}_n $}
is 
a generating function of the 
conserved densities; 
the first four 
conserved densities obtained in this 
manner 
are 
\begin{align}
 \mathcal{I}^{(0)}_{n} &= \log u_{n}, 
\nonumber \\[2mm]
 \mathcal{I}^{(1)}_{n} &= w_n -\gamma a_n b_n, 
\nonumber \\[2mm]
 \mathcal{I}^{(2)}_{n} &= 
	\frac{1}{2} \left( w_n -\gamma a_n b_n \right)^2 - \left( \gamma + \delta \right) \alpha u_{n+1}^\gamma a_n b_{n+1} 
	+ \alpha \beta u_{n+1}^{\gamma + \delta}, 
\nonumber \\[2mm]
 \mathcal{I}^{(3)}_{n} &= 
	\left[ -\left( \gamma + \delta \right) \alpha u_{n+1}^\gamma a_n b_{n+1} + \alpha \beta u_{n+1}^{\gamma + \delta} \right] 
	\left( w_n -\gamma a_n b_n 
	+ w_{n+1} - \gamma a_{n+1} b_{n+1} \right) 
\nonumber \\
& \hphantom{=} \; \mbox{}-\left( \gamma + \delta \right) \alpha^2 u_{n+1}^\gamma u_{n+2}^\gamma a_n b_{n+2}
	+ \frac{1}{3} \left( w_n -\gamma a_n b_n \right)^3. 
\nonumber 
\end{align}

Thus, two infinite sets of 
conservation laws 
for 
the generalized Toda hierarchy 
can be obtained. 
We 
consider 
linear 
combinations 
of the conserved densities
as
\begin{subequations}
\begin{align}
& \frac{1}{\gamma+\delta} \left( I^{(1)}_{n} - \mathcal{I}^{(1)}_{n} \right) = a_n b_n , 
\label{H_0'}
\\[2mm]
& \frac{1}{\gamma+\delta} \left( \gamma I^{(1)}_{n} + \delta \mathcal{I}^{(1)}_{n} \right) = w_n, 
\nonumber \\[2mm]
& \frac{1}{\gamma+\delta} \left( I^{(2)}_{n} - \mathcal{I}^{(2)}_{n} \right) \equiv w_n a_n b_n
	-\frac{1}{2} \left( \gamma - \delta \right) \left( a_n b_n \right)^2 
	+ \alpha u_n^\gamma a_{n-1}b_n + \beta u_n^\delta a_n b_{n-1}, 
\label{H_1'} \\[2mm]
& \frac{1}{\gamma+\delta} \left( \gamma I^{(2)}_{n} +\delta \mathcal{I}^{(2)}_{n} \right) \equiv \frac{1}{2} w_n^2 
	+\frac{1}{2} \gamma \delta \left( a_n b_n \right)^2 
	- \alpha \delta u_n^\gamma a_{n-1}b_n + \beta \gamma u_n^\delta a_n b_{n-1} + \alpha \beta u_n^{\gamma + \delta}, 
\label{H_2'} \\[2mm]
& \frac{1}{\gamma+\delta} \left( I^{(3)}_{n} - \mathcal{I}^{(3)}_{n} \right) \equiv 
w_n^2 a_n b_n - \left( \gamma - \delta \right) w_n \left( a_n b_n \right)^2 + \frac{1}{3} \left( \gamma^2 - \gamma \delta 
	+ \delta^2 \right) \left( a_n b_n \right)^3
\nonumber \\
& 
\hspace{15mm}
\mbox{} +\alpha u_{n+1}^\gamma a_n b_{n+1} 
	\left( w_n - \gamma a_n b_n + w_{n+1} - \gamma a_{n+1} b_{n+1} \right) 
\nonumber \\
& 
\hspace{15mm}
\mbox{} +\beta u_{n}^\delta a_n b_{n-1} 
	\left( w_{n-1} + \delta a_{n-1} b_{n-1} + w_{n} + \delta a_{n} b_{n} \right) 
\nonumber \\
& 
\hspace{15mm}
\mbox{} + \alpha^2 u_{n+1}^\gamma u_{n+2}^\gamma a_n b_{n+2} + \beta^2 u_{n}^\delta u_{n-1}^\delta a_n b_{n-2}
+ \alpha \beta u_{n}^{\gamma + \delta} \left( a_{n-1} b_{n-1}  + a_{n} b_{n} \right), 
\label{H_3'} 
\\[2mm]
& \frac{1}{\gamma+\delta} \left( \gamma I^{(3)}_{n} +\delta \mathcal{I}^{(3)}_{n} \right) \equiv
\frac{1}{3} w_n^3 + \gamma \delta w_n \left( a_n b_n \right)^2 - \frac{1}{3} \gamma \delta \left( \gamma - \delta 
	\right) \left( a_n b_n \right)^3
\nonumber \\
& 
\hspace{15mm}
\mbox{} - \alpha \delta u_{n+1}^\gamma a_n b_{n+1} 
	\left( w_n - \gamma a_n b_n + w_{n+1} - \gamma a_{n+1} b_{n+1} \right) 
\nonumber \\
& 
\hspace{15mm}
\mbox{} + \beta \gamma u_{n}^\delta a_n b_{n-1} 
	\left( w_{n-1} + \delta a_{n-1} b_{n-1} + w_{n} + \delta a_{n} b_{n} \right) 
\nonumber \\
& 
\hspace{15mm}
\mbox{} - \alpha^2 \delta u_{n+1}^\gamma u_{n+2}^\gamma a_n b_{n+2} + \beta^2 \gamma u_{n}^\delta u_{n-1}^\delta a_n b_{n-2}
+ \alpha \beta u_{n}^{\gamma + \delta} \left( w_{n-1} + w_{n} \right). 
\label{H_4'} 
\end{align}
\end{subequations}
Here, 
the symbol `$\equiv$' 
indicates 
equivalence up to a total difference, i.e., 
\mbox{$f_n \equiv g_n$} means that there 
exists 
a local function $h_n$ 
such that \mbox{$f_n -g_n = \boldsymbol{\Delta}_n^+ h_n$}; 
thus, \mbox{$f_n \equiv g_n$} implies 
\mbox{$\sum_n f_n = \sum_n g_n$} under 
appropriate 
boundary conditions. 
%

In the special case \mbox{$\gamma=\delta$},  
(\ref{H_0'}), (\ref{H_1'}) and (\ref{H_2'}) 
correspond to the Hamiltonian densities in (\ref{H_012}), 
and (\ref{H_3'}) and (\ref{H_4'}) with the aid of the Poisson 
bracket relations 
(\ref{gPB}) 
generate 
the third and fourth flows of the generalized Toda hierarchy. 
The third and fourth flows in the 
case \mbox{$\gamma \neq \delta$} can be obtained by 
applying 
a 
transformation 
like the one 
considered in subsection~\ref{subsec_gauge}.

%
%

\section{Concluding remarks}

In this paper, we proposed a new integrable generalization of the Toda lattice hierarchy, 
which can be regarded 
as a discrete analog of the generalization of the KdV hierarchy 
to the 
Yajima--Oikawa hierarchy~\cite{YO76,Cheng92}. 
The generalized Toda hierarchy admits a Lax-pair representation 
and 
two infinite sets of 
conservation laws. 
The spatial part of the Lax-pair representation for 
the generalized Toda hierarchy 
is given by (\ref{gJacobi}), 
which provides 
an interesting 
extension of the 
eigenvalue problem for 
the Jacobi operator. 
The generalized Toda hierarchy 
involves 
arbitrary 
parameters 
$\alpha$, $\beta$, $\gamma$ and $\delta$ 
and 
newly introduced dependent variables $a_n$ and $b_n$; 
with a suitable choice of the parameters, 
the generalized Toda hierarchy possesses 
a simple Hamiltonian structure 
(cf.~the Poisson 
bracket relations 
(\ref{gPB})) 
and $a_n$ and $b_n$ can be related 
by the complex conjugation reduction. 
Then, the 
first flow of the generalized Toda hierarchy 
(cf.~(\ref{reduced_first_flow})) 
can be reduced to the Yajima--Oikawa system in a continuous limit, 
while the second flow 
(cf.~(\ref{reduced_second_flow})) 
provides 
a generalization of the Toda lattice 
in Flaschka--Manakov 
variables (\ref{Toda-F}); 
it is also possible to consider the general case where 
the newly introduced dependent variables
are vector-valued functions 
(cf.~(\ref{reduced_first_flow2}) and (\ref{reduced_second_flow2})). 
Note 
that the 
generalized Toda hierarchy 
is 
different from, and apparently 
has no relation to, 
the discrete Yajima--Oikawa hierarchy 
recently studied in~\cite{Maruno16,Tsuchida18-1} 
(also see~\cite{Yu15} for 
a \mbox{$(2+1)$}-dimensional 
version~\cite{Mel83,Zakh,Nizh82} 
of the discrete Yajima--Oikawa system). 



For some special choices of the parameters, 
(a suitable linear combination of) the first and second 
flows of the generalized Toda hierarchy 
can be reduced to 
the elementary flows of 
two discrete 
nonlinear Schr\"odinger hierarchies:\ 
the Ablowitz--Ladik 
hierarchy~\cite{AL1} 
and the Konopelchenko--Chudnovsky hierarchy~\cite{Kono82,Chud1,Chud2}. 
For 
another special choice of the parameters, 
the first and second 
flows of 
the generalized Toda hierarchy 
can be linearized by a change of variables. 


 

\addcontentsline{toc}{section}{References}
\begin{thebibliography}{99}

\bibitem{Toda1}
M.\ Toda: {\em Vibration of a chain with nonlinear interaction}, 
J.\ Phys.\ Soc.\ Jpn.\ {\bf 22} (1967) 431--436.

\bibitem{Toda2}
M.\ Toda: {\em Theory of Nonlinear Lattices},  
2nd enlarged edition
(Springer Series in Solid-State Sciences {\bf 20}, Berlin, 1989).

\bibitem{Flaschka1}
H.\ Flaschka: 
{\em The Toda lattice.\ I.\ Existence of integrals}, 
Phys.\ Rev.\ B {\bf 9} (1974) 1924--1925.

\bibitem{Flaschka2}
H.\ Flaschka:
{\em On the Toda lattice.\ II Inverse-scattering solution},
Prog.\ Theor.\ Phys.\
{\bf 51} (1974) 703--716.

\bibitem{Manakov74}
S.\ V.\ Manakov: {\em Complete integrability and 
stochastization of discrete dynamical systems}, 
Sov.\ Phys.\ JETP {\bf 40} (1975)
269--274.

\bibitem{Suris03}
Y.\ B.\ Suris: 
{\it The Problem of Integrable Discretization:\ Hamiltonian Approach}
(Birkh\"auser, Basel, 2003). 

\bibitem{Henon74}
M.\ H\'enon: {\em Integrals of the Toda lattice}, 
Phys.\ Rev.\ B {\bf 9} (1974) 1921--1923.

\bibitem{Lax}
P.\ D.\ Lax: {\em Integrals of nonlinear equations of evolution and
  solitary waves}, Commun.\ Pure Appl.\ Math.\ {\bf 21} (1968) 467--490.

\bibitem{YO76}
N.\ Yajima and M.\ Oikawa: 
{\em Formation and interaction of sonic-Langmuir solitons 
---Inverse scattering method---}, 
Prog.\ Theor.\ Phys.\ {\bf 56} (1976) 1719--1739.

\bibitem{Cheng92}
Y.\ Cheng: {\em Constraints of the Kadomtsev--Petviashvili hierarchy}, 
J.\ Math.\ Phys.\ {\bf 33} (1992) 3774--3782. 

\bibitem{Maruno16}
J.\ Chen, Y.\ Chen, B.-F.\ Feng, K.\ Maruno and Y.\ Ohta: 
{\em An integrable semi-discretization of the coupled Yajima--Oikawa system}, 
J.\ Phys.\ A:\ Math.\ Theor.\ {\bf 49} (2016) 165201, 
arXiv:1509.06996 [nlin.SI]. 

\bibitem{Yu15}
G.-F.\ Yu and Z.-W.\ Xu: 
{\em Dynamics of a differential-difference integrable $(2+1)$-dimensional system}, 
Phys.\ Rev.\ E {\bf 91} (2015) 062902.

\bibitem{Tsuchida18-1}
T.\ Tsuchida: 
{\em Integrability of a discrete Yajima--Oikawa system}, 
 arXiv:1804.10224 [nlin.SI] (2018). 

\bibitem{AL1}
M.\ J.\ Ablowitz and J.\ F.\ Ladik: 
{\em Nonlinear differential--difference equations and Fourier analysis}, 
J.\ Math.\ Phys.\ {\bf 17} (1976) 1011--1018. 



\bibitem{Kono82}
B.\ G.\ Konopelchenko:
{\em Elementary B\"acklund transformations, nonlinear superposition 
principle and solutions of the integrable equations}, 
Phys.\ Lett.\ A {\bf 87} (1982) 445--448.

\bibitem{Chud1}
D.\ V.\ Chudnovsky and G.\ V.\ Chudnovsky: 
{\em B\"acklund transformation as a method of decomposition and reproduction 
of two-dimensional nonlinear systems}, 
Phys.\ Lett.\ A {\bf 87} (1982) 325--329.

\bibitem{Chud2}
D.\ V.\ Chudnovsky and G.\ V.\ Chudnovsky: 
{\em B\"acklund transformations and lattice systems with G-gauge symmetries}, 
Phys.\ Lett.\ A {\bf 89} (1982) 117--122.

\bibitem{YO2}
N.\ Yajima and M.\ Oikawa: 
{\em A class of exactly solvable nonlinear evolution equations}, 
Prog.\ Theor.\ Phys.\ {\bf 54} (1975) 1576--1577. 

\bibitem{Ab78}
M.\ J.\ Ablowitz: 
{\em Lectures on the inverse scattering transform},
Stud.\ Appl.\ Math.\ {\bf 58} (1978) 17--94.

\bibitem{New79}
A.\ C.\ Newell: 
{\em The general structure of integrable evolution equations}, 
Proc.\ R.\ Soc.\ Lond.\ A {\bf 365} (1979) 283--311.


\bibitem{Mak82}
V.\ G.\ Makhan'kov and O.\ K.\ Pashaev: 
{\em Nonlinear Schr\"{o}dinger 
equation with noncompact isogroup}, 
Theor.\ Math.\ Phys.\ {\bf 53} (1982) 979--987. 



\bibitem{YCMa81}
Y.-C.\ Ma: 
{\em The resonant interaction among long and short waves}, 
Wave Motion {\bf 3} (1981) 257--267. 

\bibitem{Mel83}
V.\ K.\ Mel'nikov: 
{\em On equations for wave interactions}, 
Lett.\ Math.\ Phys.\ {\bf 7} (1983) 129--136.

\bibitem{MPK83}
V.\ G.\ Makhankov, O.\ K.\ Pashaev and A.\ Kundu: 
{\em Integrable reductions of manycomponent magnetic systems in $(1, 1)$ dimensions}, 
Physica Scripta {\bf 28} (1983) 229--234.

\bibitem{Dub88}
B.~A.~Dubrovin, T.~M.~Malanyuk, I.~M.~Krichever 
and V.~G.~Makhan'kov: 
{\em Exact solutions of the time-dependent Schr\"odinger equation 
with self-consistent potentials}, Sov.\ J.\ Part.\ Nucl.\ {\bf 19} (1988) 252--269, 
http://people.sissa.it/\~{}dubrovin/bd\_{}papers.html .

\bibitem{Li2016}
C.-Z. Li: 
{\em Constrained lattice-field hierarchies and Toda system with Block symmetry}, 
Int.\ J.\ Geom.\ Methods Mod.\ Phys.\ {\bf 13} (2016) 1650061, 
arXiv:1602.07145 [nlin.SI].  

\bibitem{GI82}
V.\ S.\ Gerdzhikov and  M.\ I.\ Ivanov:
{\em Hamiltonian structure of multicomponent nonlinear
Schr\"{o}dinger equations in difference form},
Theor.\ Math.\ Phys.\ {\bf 52} (1982) 676--685.


\bibitem{Zhang91}
H. Zhang,  G.-Z.\
Tu, W.\ Oevel 
and B.\ Fuchssteiner: 
{\em Symmetries, conserved 
quantities, and hierarchies for some lattice systems with soliton structure}, 
J.\ Math.\ Phys.\ {\bf 32} (1991) 1908--1918.




\bibitem{Tu}
I.\ Merola, O.\ Ragnisco and G.-Z.\ 
Tu: 
{\em A novel hierarchy of integrable lattices},  
Inverse Probl.\ {\bf 10} (1994) 1315--1334.

\bibitem{Haberman}
R.\ Haberman: 
{\em An infinite number of conservation laws for coupled nonlinear 
evolution equations}, J.\ Math.\ Phys.\ {\bf 18} (1977) 1137--1139.

\bibitem{Wadati1977}
M.\ Wadati and M.\ Watanabe: 
{\em Conservation laws of a Volterra system and nonlinear self-dual network equation}, 
Prog.\ Theor.\ Phys.\ {\bf 57} (1977) 808--811.

\bibitem{Zakh}
V.\ E.\ Zakharov: 
{\em The inverse scattering method}, 
``Solitons'' edited by R.~K.~Bullough and P.~J.~Caudrey 
(Topics in Current Physics 17,
Springer, Berlin, 1980) pp.\ 243--285. 


\bibitem{Nizh82}
L.\ P.\ Nizhnik and M.\ D.\ Pochinaiko:
{\em Integration of the nonlinear two-dimensional spatial 
Schr\"odinger equation by the inverse-problem method},
Funct.\ Anal.\ Appl.\ {\bf 16} (1982) 66--69.


\end{thebibliography}
\end{document}